# Acoustic Signatures of Pinch-Off Cavities During Water-Entry

Zirui Liu,[1,2] Tongtong Ding,[1] Mingyue Kuang,[1] Zimeng Li,[1] Junyi Zhao,[3] A-Man Zhang,[1,2,3,4,a] and Shuai Li[1,2,3,4,b]

[1] *College of Shipbuilding Engineering, Harbin Engineering University, Harbin, 150001, China*

[2] *Nanhai Institute of Harbin Engineering University, Sanya, 572024, China*

[3] *Southampton Ocean Engineering Joint Institute, Harbin Engineering University, Harbin, 150001, China*

[4] *National Key Lab of Ship Structural Safety, Harbin Engineering University, Harbin, 150001, China*

**Abstract:** This study experimentally, numerically, and theoretically investigates the cavity/bubble dynamics and radiated acoustics during the water entry of a centimeter-scale cylindrical projectile with a conical nose. Experiments were conducted in a laboratory tank, employing synchronized high-speed imaging and hydrophone measurements to characterize the cavity closure modes and their resultant acoustic signatures across a range of Froude numbers. The acoustic signal features a weak radiated signal upon impact, followed by significant pressure oscillations spanning more than 20 cycles in the flow field after cavity elongation and pinch-off. A numerical model based on the Finite Volume

---

[a] zhangaman@hrbeu.edu.cn

[b] lishuai@hrbeu.edu.cn

Method (FVM) successfully captures these physical processes. Subsequently, a semi-theoretical model that incorporates the projectile's boundary effect is developed from potential flow theory. The model not only yields a dominant cavity oscillation frequency that agrees well with experimental data, but also reveals that the boundary effect leads to a cavity oscillation frequency markedly higher than the Minnaert frequency of an equivalent-volume ellipsoidal bubble containing an internal rigid core. The dominant cavity frequency falls nearly linearly with $Fr$, governed by nose geometry and projectile inertia. This study clarifies the underlying physics connecting cavity dynamics during water entry to underwater acoustic radiation.

**Keywords**：water entry; bubble dynamics; underwater sound; acoustic signal

## I. INTRODUCTION

As a classic problem in fluid mechanics, water entry has attracted enduring interest owing to its scientific importance and relevance to both natural phenomena (Bush et al., 2006; Jung, 2025) and technological applications (Xie et al., 2018). As a solid penetrates a free surface at high speed, it transfers sufficient kinetic energy to the fluid. causing violent deformation and forming a transient cavity. The cavity subsequently expands and collapses under the combined influence of inertia and hydrostatic pressure. This process produces complex multiphase phenomena such as free-surface splashing (Eshraghi et al., 2020), jet formation (Gekle et al., 2009) and acoustic radiation (Abelson, 1970). A deeper understanding of this dynamic process not only clarifies the fundamental physics of transient fluid–structure interactions but also, as demonstrated in studies on underwater vehicle noise prediction (Özden et al., 2016) and acoustic positioning system design (Chen et al., 2002), has significant implications for related engineering applications.

A considerable body of experimental and theoretical research has focused on cavity evolution during water entry, revealing a rich variety of flow features. Closure modes are typically categorized into two types (Aristoff et al., 2009): (i) deep seal, characterized by midsection cavity collapse and pinch-off driven by hydrostatic pressure (Bergmann et al., 2009), and (ii) surface seal, in which the splash curtain closes at the free surface due to pressure differences and surface tension (Eshraghi et al., 2020). At pinch-off in a deep seal, two distinct jets emerge from the collapse singularity: a downward jet and an upward one that pierces the free surface, known as the classical Worthington jet (Gekle et al., 2010). Research has primarily focused on the pinch-off time and location in deep seals, leading to substantial theoretical progress. Results indicate that the cavity depth and pinch-off location are governed by the projectile diameter and scale with the Froude number according to a power-law relation (Glasheen et al., 1996; Shi et al., 2019). The pinch-off location is influenced by the radial expansion and impact velocities during cavity formation. Duclaux et al. (2007) compared cavity dynamics from formation to collapse for spheres and cylinders. They proposed an analytical model based on the Besant–Rayleigh equation and potential-flow theory, showing that cavity characteristics depend on both object geometry and impact velocity. Subsequent studies have shown that the effective mass and density, which are determined by the projectile's geometry (Kim et al., 2019) and material properties (Aristoff et al., 2009), modify the inertial effects during water entry and consequently the scaling of cavity pinch-off. Under certain conditions (Gilbarg et al., 1948; Aristoff et al., 2009; Hong et al., 2023), the cavity closure mode transitions progressively from deep to surface seal. During surface seal, the cavity becomes isolated from the atmosphere, which alters the pinch-off characteristics (Eshraghi et al., 2020; Mansoor et al., 2014). Consequently, the closure mechanism plays a crucial role in dictating the subsequent cavity evolution (Zhang et al., 2022). These complex cavity dynamics, especially the closure process, are closely linked to the resulting underwater acoustic signatures, a topic that has garnered increasing research interest.

With advances in experimental techniques, acoustic emissions generated during water entry have become a major research focus (Sun et al., 2021; Li et al., 2021; Nelli et al., 2024). Grumstrup et al. (2007) experimentally observed stationary ripples forming on the surface of the attached cavity during high-speed water entry of a sphere. They found that the product of the ripple wavelength ($\lambda$) and the cavity oscillation frequency ($f$) approximately equals the sphere's impact velocity ($U_0$), i.e., $U_0 \approx \lambda f$. This behavior arises from acoustic oscillations excited by the deep seal closure, with a frequency close to the Minnaert frequency of an equivalent spherical bubble (Grumstrup et al., 2007; Ueda et al., 2022). Yu et al. (2023) further confirmed experimentally that the ripples are stationary. They attributed the acoustic source to radial vibration at the leading edge of the cavity, whose natural frequency governs both the ripple formation and the far-field acoustic response. For cylindrical bodies, the cavity oscillation frequency is modified by the solid volume enclosed within the cavity and by the geometric constraints on radial displacement at the contact interface. Through theoretical analysis and coupled numerical simulations, Zhang et al. (2024b) modeled the cavity as a hollow cylindrical structure and elucidated the influence of rod radius on the ripple frequency during constant-velocity entry. In summary, while existing research has predominantly examined the acoustic characteristics of cavity ripples following pinch-off under deep-seal conditions, the influence of different closure modes on pre–pinch-off acoustic emissions remains unclear. In addition, projectile geometry modifies the distribution of hydrodynamic added mass and the surrounding pressure field, thereby affecting the inertial forces and the projectile's velocity history during water entry. To systematically decouple these effects, we conducted a controlled water-entry experiment using projectiles of different geometric sizes. The corresponding experimental setup is schematically illustrated in FIG 1. Acoustic signals were captured by a single hydrophone synchronized with high-speed imaging to correlate the acoustic signatures with the cavity dynamics. Based on this experimental approach, our aim is to determine the acoustic characteristics of cavities under different closure modes, and to clarify the respective roles of projectile geometry and inertial effects in shaping the temporal and spectral features of the cavity acoustic signals.

The structure of this paper is organized as follows. Section II provides detailed descriptions of the experimental setup and the numerical methods. Section III first analyzes the cavity closure modes and the corresponding acoustic responses across different Froude numbers, based on combined experimental and numerical results. It then develops a semi-theoretical model for cavity oscillation that incorporates the projectile's boundary effect within the framework of potential flow theory, followed by an examination of the influence of the Froude number and projectile geometry on the oscillation frequency. Finally, Section IV summarizes the main conclusions.

## II. METHODOLOGY

### A. Experimental setup

As shown in FIG 1, the water-entry experiments were conducted in a laboratory tank measuring 500 mm × 500 mm × 500 mm, with a wall thickness of 8 mm, at an ambient temperature of approximately 20 °C and atmospheric pressure (101 kPa). The projectile is a cylinder with a 90° conical nose, made of stainless steel. To investigate the effects of projectile inertia and volume, the diameter was fixed at $D = 10$ mm for all models to ensure an identical wetted surface area of the nose. Three projectiles with different aspect ratios, denoted by $L^* = L/D$ (where $L$ is the total length of the projectile), were tested. TABLE I presents the physical quantities and dimensionless parameters considered in this study, along with their respective ranges. Since the tank width was much greater than the projectile diameter, sidewall effects on the water-entry cavity were negligible (Mansoor et al., 2014). The water depth was approximately 400 mm, and the maximum cavity depth prior to pinch-off was about 200 mm, ensuring that bottom-wall effects were also negligible.

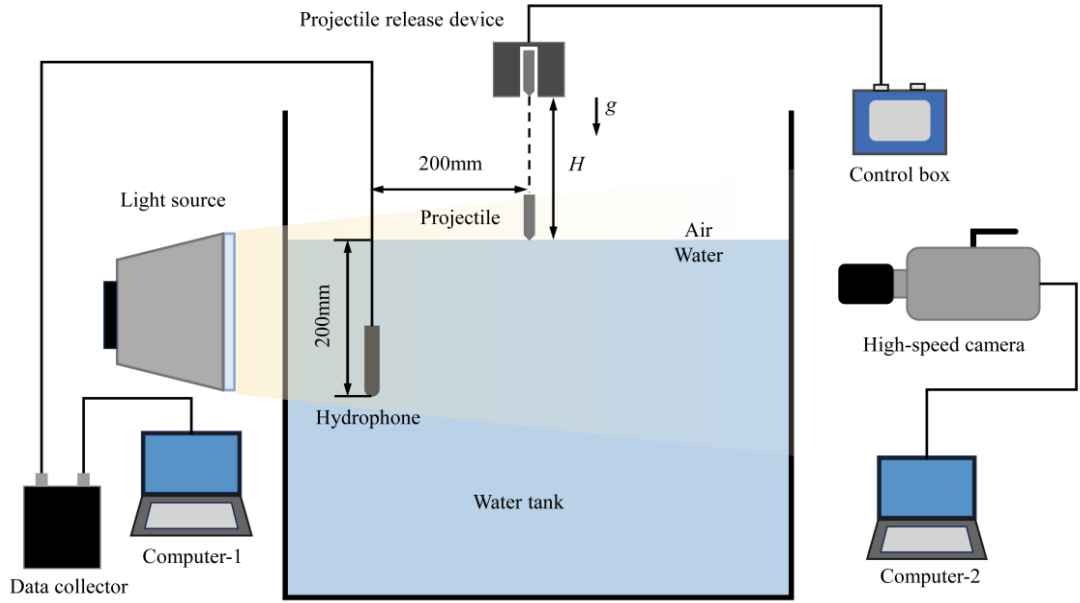

FIG 1. (Color online) Schematic diagram of the experimental setup for water-entry acoustic signal measurement.

A Phantom V2012 high-speed camera (USA) with a macro lens was employed to capture the projectile motion and cavity evolution during water entry, at a resolution of 800 × 1280 pixels and a frame rate of 5000 fps. Acoustic signals were recorded using a Brüel & Kjær Type 8106 hydrophone connected to a matching data acquisition system (Denmark). The hydrophone enables absolute measurements of sound pressure across a frequency range of 7 Hz to 80 kHz, with a receiving sensitivity of −173 ± 3 dB re 1 V/μPa. Its directivity pattern is approximately ± 2 dB re 1 V/μPa in both the horizontal and vertical planes, characterized at 100 kHz. The hydrophone, operated at a sampling frequency of 6.4 kHz, was positioned 0.2 m below the free surface and 0.2 m horizontally from the impact point. The projectile release mechanism was mounted directly above the tank center and fixed to a height-adjustable linear slide, enabling precise control of the water-entry velocity. The theoretical impact velocity is given by $U_0 = \sqrt{2gH}$, where $g$ = 9.81 m/s² is the gravitational acceleration and $H$ is the release height. The actual impact velocity was determined from high-speed image analysis.

TABLE I. Projectile characteristics and relationships of dimensionless numbers

| Parameters | Symbols/definition | Value/ranges | Units |
| --- | --- | --- | --- |
| Projectile diameter | $D$ | 10 | mm |
| Aspect Ratio | $L^* = L/D$ | 2,3,4 | — |
| Projectile mass | $m$ | 10.2,16.3,21.5 | g |
| Release height | $H$ | 0.15~0.63 | m |
| Impact velocity | $U_0 = \sqrt{2gH}$ | 1.7~3.5 | m/s |
| Reynolds number | $Re = U_0 D/\nu$ | $1.7\times10^4$~$3.5\times10^4$ | — |
| Froude number | $Fr = U_0/\sqrt{gD}$ | 5.4~11.2 | — |

**B. Numerical setup**

The strong axisymmetry of the flow observed during water entry justified the use of a two-dimensional axisymmetric computational domain to improve simulation efficiency. The computational model, with the symmetry axis aligned along the projectile centerline, is shown in FIG 2(a). The domain was sufficiently large to minimize boundary effects on cavity evolution, with a rectangular extent of $70D \times 50D$ and a water depth of $50D$. Boundary conditions included a pressure-outlet top boundary set to atmospheric pressure and no-slip conditions on all other walls. The projectile geometry was identical to that used in the experiments, and the pressure monitoring point was positioned to coincide with the hydrophone location.

The overset grid technique was employed to simulate the projectile motion. As shown in FIG 2(b), the computational mesh was divided into a background domain and an overset domain, with the latter assigned a single degree-of-freedom motion. To balance the accuracy of cavity interface capturing with computational efficiency, grid refinement was applied solely to the area around the projectile trajectory and the free surface, with a minimum cell size of $\Delta x = 0.0375D$. The mesh was gradually coarsened toward the outer regions, while the overset mesh shared the same resolution as the refined zone. A graded prismatic mesh was generated along the projectile surface to accurately resolve the near-wall flow. An adaptive time-stepping scheme was implemented to maintain a maximum Courant number (CFL) below 0.4.

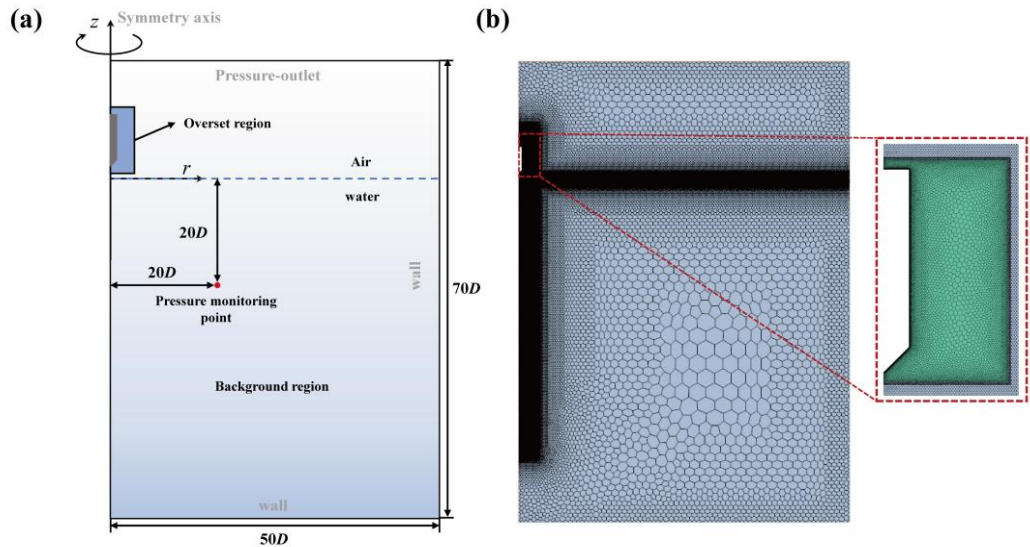

FIG 2. (Color online) (a) Schematic diagram of the computing domain and setting of boundary conditions; (b) Computational domain meshing: background region and overset region.

The numerical simulations employed the finite volume method (FVM) to solve the Navier–Stokes (NS) equations for a gas–liquid two-phase flow, while neglecting heat transfer and natural cavitation effects. The Volume of Fluid (VoF) method (Hirt et al., 1981) was adopted to capture the gas–liquid interface. Spatial and temporal discretizations employed a bounded central differencing scheme and a first-order implicit Euler method, respectively. The pressure–velocity coupling was performed using the SIMPLE algorithm (Patankar et al., 1983). The Reynolds-Averaged Navier–Stokes (RANS) equations, combined with the Shear Stress Transport (SST) model, were used to account for turbulence (Menter et al., 1994). Under these conditions, the governing equations can be expressed as follows:

$$\frac{\partial \varrho_m}{\partial t} + \nabla \cdot (\varrho_m \boldsymbol{u}) = 0, \tag{1}$$

$$\frac{\partial (\varrho_m \boldsymbol{u})}{\partial t} + \nabla \cdot (\varrho_m \boldsymbol{u} \otimes \boldsymbol{u}) = -\nabla \cdot pI + \nabla \cdot (\overline{T} + T_{RNAS}) + \boldsymbol{F}, \tag{2}$$

where $\varrho_m$ denotes the mixture density of the fluid, $\boldsymbol{u}$ the velocity vector, $p$ the fluid pressure, $\overline{T}$ the mean viscous stress tensor, $T_{RNAS}$ the Reynolds stress tensor, and $\boldsymbol{F}$ the resultant body force. The viscous stress tensor incorporates both laminar and turbulent contributions through the effective viscosity. Turbulence closure is provided by the SST model, whose transport equations read:

$$\frac{\partial \varrho_m k}{\partial t} + \nabla \cdot (\varrho_m k \boldsymbol{u}) = \nabla \cdot \left[ (\mu_m + \sigma_k \mu_t) \nabla k \right] - f_{\beta^*} \beta^* \varrho_m \omega k + P_k, \tag{3}$$

$$\frac{\partial \varrho_m \omega}{\partial t} + \nabla \cdot (\varrho_m \omega \boldsymbol{u}) = \nabla \cdot \left[ (\mu_m + \sigma_\omega \mu_t) \nabla \omega \right] - f_\beta \beta \varrho_m \omega^2 + P_\omega, \tag{4}$$

where $k$ is the turbulent kinetic energy, $\mu_m$ is the dynamic viscosity of the mixture, $\mu_t$ is the turbulent viscosity, $\omega$ is the turbulence frequency, $\sigma_k$ and $\sigma_\omega$ are turbulence model constants, $f_\beta$ is the vortex stretching modification factor, $f_{\beta^*}$ is the free-shear modification factor, the model coefficients are denoted by $\beta$ and $\beta^*$, and the production terms by $P_k$ and $P_\omega$.

The VOF model employs a single-fluid formulation, where all phases share a common pressure and velocity field under the no-slip condition. The interface is tracked using phase volume fractions, with the local composition determined by the volume fraction of each phase within a computational cell. The volume fractions of water and air, denoted by $a_{\text{water}}$ and $a_{\text{air}}$, satisfy the following constraint:

$$a_{\text{water}} + a_{\text{air}} = 1, \tag{5}$$

In the numerical simulation, $a_{\text{water}}$ and $a_{\text{air}}$ denote the volume fractions of water and air within each computational cell, respectively. Cells with $0 < a < 1$ represent the gas–liquid interface. Water is modeled as an incompressible fluid with a constant density of 1000 kg/m³. To capture the periodic pulsation of the cavity, the air phase is treated as a compressible, adiabatic ideal gas (Zhang et al., 2024b). Its Equation of State (EOS) is given by:

$$\varrho_{\text{air}} = \varrho_{\text{air},0} \left( \frac{P}{P_0} \right)^{\frac{1}{\kappa}}, \tag{6}$$

where $P_0$ is the standard atmospheric pressure, $\rho_{\text{air},0}$ is the air density at standard atmospheric pressure, taken as 1.28 kg/m³, and $\kappa$ is the adiabatic index for an ideal gas, taken as 1.4.

## III. RESULTS AND DISCUSSION

### A. The evolution of cavity morphology

With the moment of water-entry impact defined as the initial time ($t = 0$), the cavity evolution is compared for four different water-entry velocities. FIG 3 presents experimental images of the cavity dynamics for the projectile with $L^* = 4$ under different initial impact velocities. Upon impact, a splash forms above the free surface. As the projectile descends and displaces the surrounding liquid, it generates a cavity with an outwardly expanding splash crown. By $t = 20$ ms, differences in cavity depth become apparent: a larger $Fr$ corresponds to higher initial kinetic energy and thus a deeper cavity. As the projectile continues to move downward, the cavity keeps expanding, and its volume increases. By $t = 30$ ms, the splash crown begins to contract radially toward the center, with stronger contraction observed for higher $Fr$.

By $t = 40$ ms, the cavity shapes exhibit significant variation among the different $Fr$. At $Fr = 8.0$, the splash crown gradually falls back without noticeable axial contraction (FIG 3(a)). At $Fr = 9.6$, in contrast, the crown closes above the cavity, sealing it from the atmosphere and cutting off further air entrainment (FIG 3(d)). The inertial expansion of the cavity is increasingly suppressed by hydrostatic pressure as it descends. This reversal of the radial flow subsequently triggers cavity collapse. Around $t \approx 54$ ms, the cavity pinches off at mid-depth in all four cases, splitting into upper and lower sub-cavities. The upper part retracts toward the free surface, forming a Worthington jet, while the lower part remains attached to the projectile. The tail cavity that follows the motion of the projectile differs significantly from the spherical bubbles commonly discussed in the literature (Zhang et al., 2024a; Jamali et al., 2025). A detailed quantitative analysis of this behavior will be presented in the following sections.

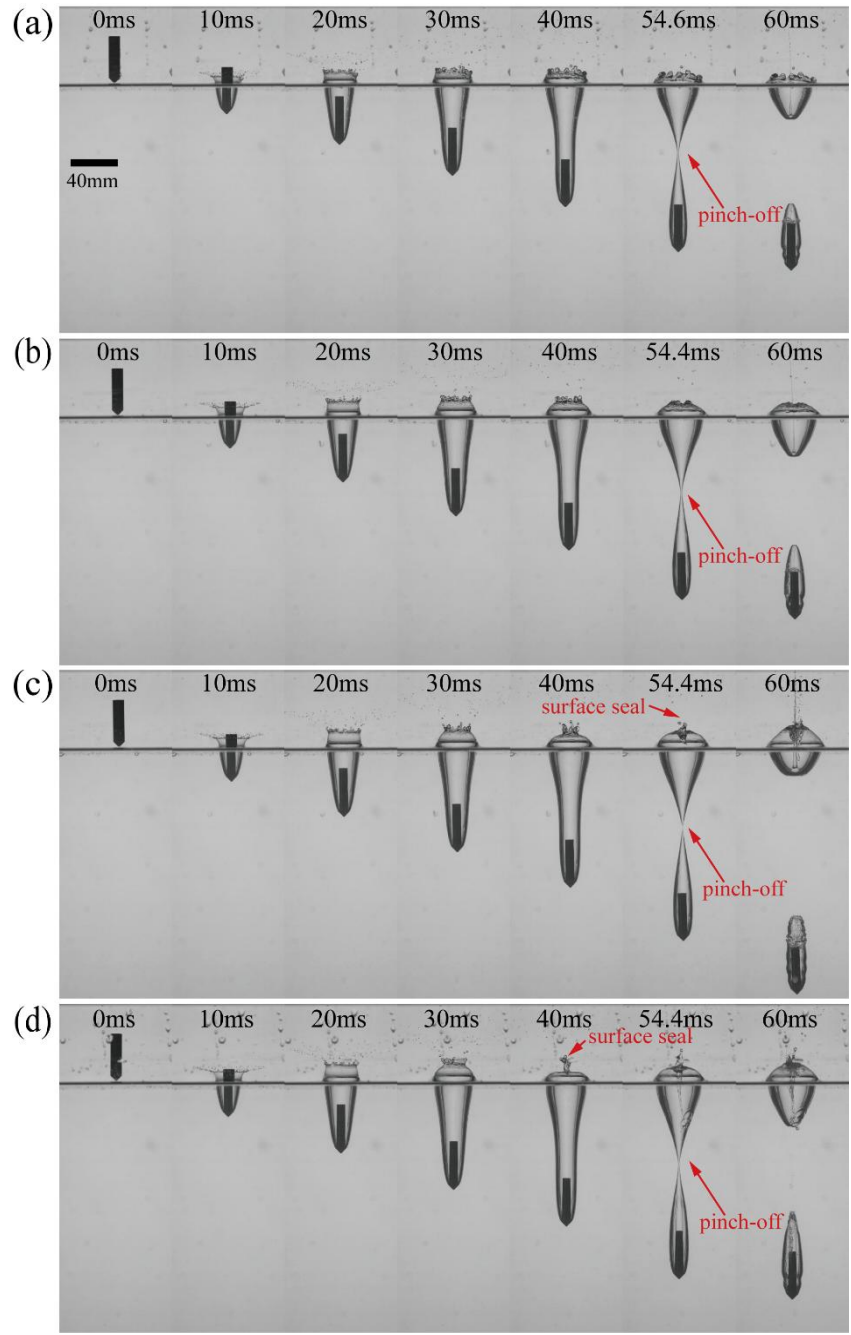

FIG 3. (Color online) Images of water-entry cavities at various impact velocities ($L^*=4$): (a) $Fr = 8.0$ ;(b) $Fr = 8.9$ ;(c) $Fr = 9.3$ ;(d) $Fr =9.6$.

It is worth noting that for $Fr = 8.0$ and $Fr = 8.9$, the splash crown near the free surface does not fully close, and the cavity remains connected to the atmosphere, corresponding to the deep-seal mode. Moreover, the timing of surface closure relative to mid-depth pinch-off depends on $Fr$. At $Fr = 9.3$, the surface closure and pinch-off occur nearly simultaneously, representing a transition from deep-seal to surface-seal modes. At $Fr = 9.6$, surface closure precedes pinch-off and induces a downward jet. During the closure process, shear and pressure disturbances arise due to cavity separation and breakup. By $t = 60$ ms, pronounced ripples appear on the cavity wall attached to the projectile in all cases, indicating that volumetric oscillations of the cavity have already been initiated.

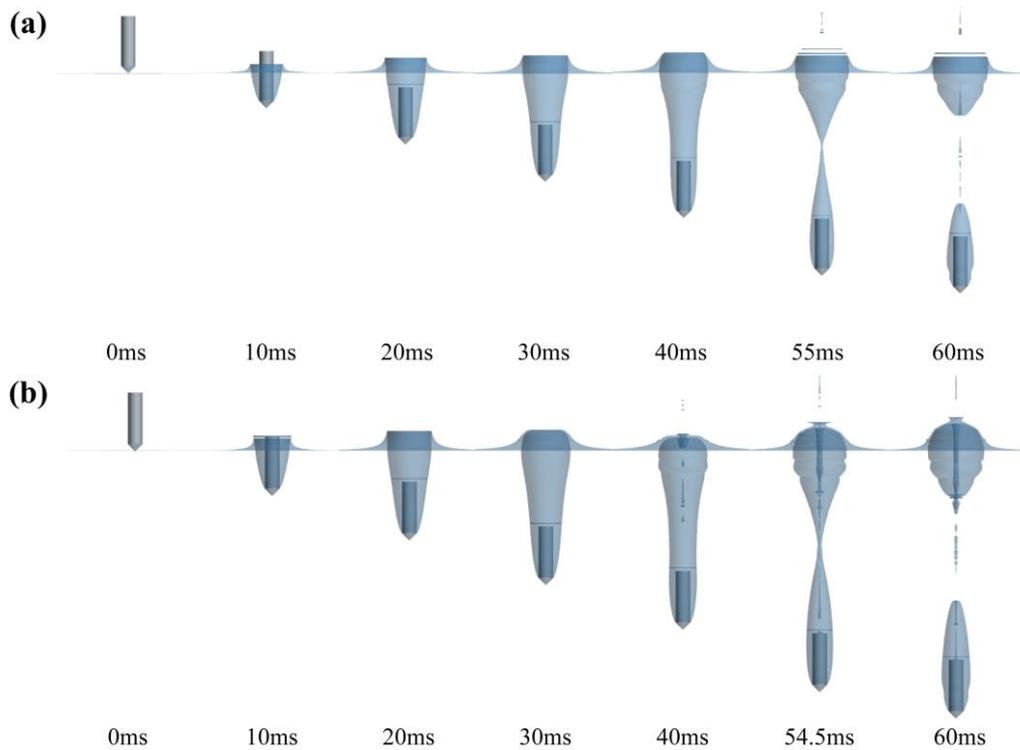

FIG 4. (Color online) Numerical simulation results of the deep seal and surface seal phenomena in the experiment shown in FIG 3 (a) and (d).

To validate the numerical model, two representative cases were simulated, with the results shown in FIG 4 (a) and (b) corresponding to the experimental cases in FIG 3(a) and (d), respectively. The simulations were initialized with the projectile's conical tip positioned flush with the free surface and initial velocities of 2.5 m/s and 3.0 m/s. Overall, the numerical results show good agreement with experiments in terms of cavity evolution and projectile trajectory. However, due to mesh resolution limitations, the simulations do not fully capture the secondary droplets detached from the splash crown, resulting in slight discrepancies in morphology. Despite this, the global cavity dynamics are accurately reproduced. Notably, at $t = 40$ ms in FIG 4(b), the simulation captures the splash crown converging along the axis to form a jet that leads to cavity closure. The predicted cavity pinch-off times agree closely with experiments, with errors within 7.2%, well within the acceptable range for cavity dynamics studies. Furthermore, the ripples on the cavity surface observed at $t = 60$ ms in the numerical results confirm that the model accurately captures the volumetric oscillations following pinch-off.

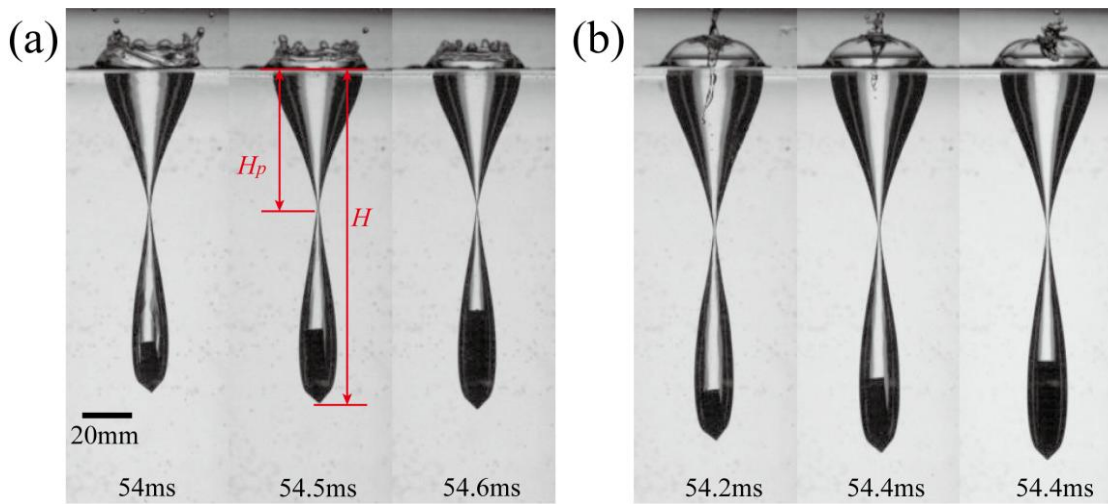

FIG 5. (Color online) Cavity morphology evolution for different projectile lengths in two seal modes ($L^* = 2$, 3 and 4 from left to right): (a) Deep seal, $Fr = 8.3$; (b) Surface

seal, $Fr = 9.3$.

Subsequently, the effect of aspect ratio on cavity morphology evolution is examined for projectiles at a fixed water-entry velocity. FIG 5 presents experimental snapshots of cavity pinch-off (both deep and surface seal) for projectiles with $L^* = 2$, 3, and 4. The projectile with $L^* = 2$ exhibits noticeable trajectory deviation, attributed to its lower initial kinetic energy and reduced ballistic stability. However, this deviation has minimal impact on the overall cavity morphology. At $Fr = 8.3$, the surface splash crown and the underwater cavity pinch-off are remarkably similar across all cases, with deep-seal closure observed. In contrast, at $Fr = 9.3$, the surface splash converges axially to form a dome, while all cavities undergo surface-seal closure first.

Table II. Measured values of $t_p$, $H_p$, $H$ and $H_p/H$ when $Fr = 8.3$ and 9.3

| $Fr$ | $L^*$ | $t_p U_0/D$ | $H_p/D$ | $H/D$ | $H_p/H$ |
|---|---|---|---|---|---|
| 8.3 | 2 | 14.09 | 5.85 | 13.65 | 0.429 |
| 8.3 | 3 | 14.28 | 5.91 | 14.13 | 0.418 |
| 8.3 | 4 | 14.21 | 5.92 | 14.41 | 0.409 |
| 9.3 | 2 | 15.75 | 6.90 | 15.80 | 0.437 |
| 9.3 | 3 | 15.92 | 6.91 | 16.22 | 0.425 |
| 9.3 | 4 | 15.78 | 6.93 | 16.50 | 0.418 |

The measured characteristic dimensions of cavity closure for the experiments shown in FIG 5 are summarized in Table II, where $t_p$ denotes the cavity pinch-off time, $H_p$ represents the pinch-off depth, and $H$ is the depth of the projectile tip at pinch-off. The pinch-off time is influenced by the projectile's $L^*$: as $L^*$ increases, the effective mass of the projectile grows (Kim et al., 2019), resulting in higher initial kinetic energy. Consequently, the velocity decays more slowly after water entry, leading to a more stable cavity and a delayed pinch-off. In contrast, the time to surface closure occurs earlier for higher $L^*$. Regarding cavity morphology, since the projectile nose geometry is identical, the radial cavity profiles remain largely consistent across cases. The primary differences appear in the axial dimension, specifically the total cavity length: both $H_p$ and $H$ increase with $L^*$, while the ratio $H_p/H$ decreases, progressively deviating from 0.5. This trend is consistent with earlier research (Aristoff et al., 2009; Kim et al., 2019), confirming that a higher aspect ratio $L^*$ leads to a larger cavity volume after pinch-off.

### B. Acoustic pressure signal characteristics

FIG 6 shows the full cavity evolution and the corresponding acoustic signal for the $L^* = 4$ projectile at $Fr = 8.0$. The timeline in (b) indicates the moments corresponding to the key cavity stages in (a). Due to the low initial velocity leading to deep-seal closure, the water-entry process can be divided into two phases: (A) from the projectile's impact on the free surface to cavity necking and pinch-off, and (B) from pinch-off to subsequent sustained cavity oscillations. Projectile impact occurred at $t_0 = 0$ ms, coinciding with the onset of variation in the acoustic signal. Following impact, the signal initially increases and gradually transitions from positive to negative. Throughout Phase A, the acoustic signal remains relatively stable without significant response, despite the continuous increase in cavity depth. At $t_4 = 54.6$ ms, the cavity undergoes radial contraction at its midsection, resulting in deep-seal closure. This event induces a sharp acoustic pulse, characterized by an instantaneous signal surge that attains its peak amplitude.

Between 54.6 ms and 120 ms, the cavity experiences volume oscillations driven by shear forces generated during pinch-off and by hydrostatic pressure. These oscillations manifest as ripples on the cavity surface and are accompanied by bubble shedding from the cavity tail. The acoustic signal gradually decays after the initial cavity closure, while subsequent bubble shedding continuously generates additional oscillatory signals. Overall, the acoustic signature during this stage is characterized by the superposition of decaying pressure waves from the deep pinch-off and disturbances arising from new bubble shedding (Zhang et al., 2020). FIG 6(c) presents the Fast Fourier Transform (FFT) of the sound pressure signal shown in FIG 6(b). The spectrum exhibits a prominent peak at approximately 360 Hz, indicating the dominant frequency of the acoustic signal. The time interval between the first two positive peaks after $t_4$ yields $\Delta t$ = 2.8 ms, corresponding to a frequency of $f = 1/\Delta t \approx 357$ Hz. This dominant frequency aligns closely with the volumetric oscillation period of the cavity during Phase B. This consistency validates that post-pinch-off volumetric oscillations are the primary mechanism driving the low-frequency components of the acoustic signature. Additionally, the spectrum exhibits a broadband distribution at high frequencies which, as noted by Sun et al. (2021) and further supported by Nelli et al. (2025), originates from random-phase oscillations of numerous microbubbles during their formation and collapse.

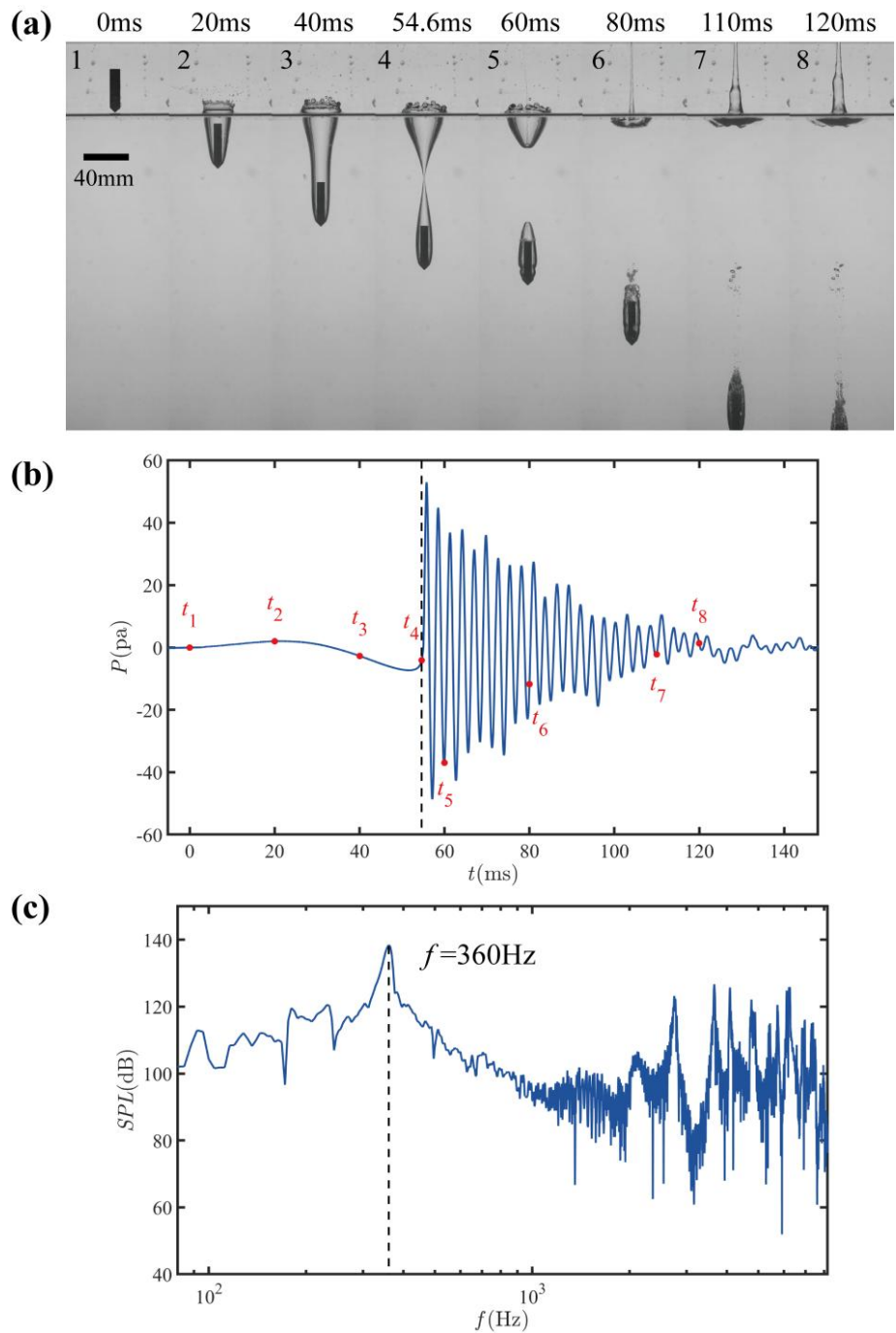

FIG 6. (Color online) Coupling characteristics of acoustic signals and cavity evolution during the water entry at $L^* = 4$ and $Fr = 8.0$: (a) Representative cavity shapes at critical moments; (b) Time-domain waveform of the sound pressure signal; (c) Frequency-domain spectrum obtained by FFT.

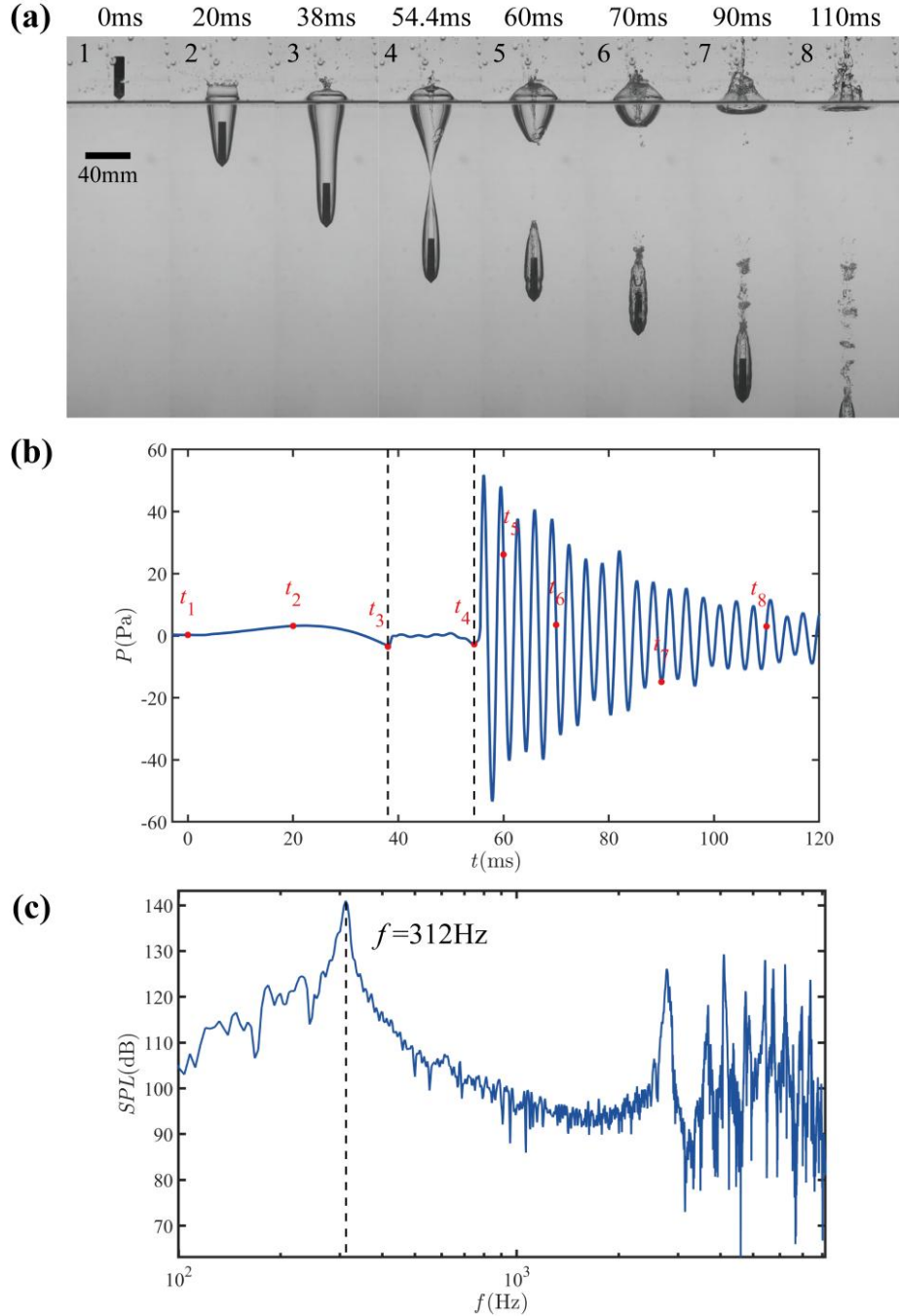

FIG 7. (Color online) Coupling characteristics of acoustic signals and cavity evolution during the water entry $L^* = 4$ and $Fr = 9.6$: (a) Representative cavity shapes at critical moments; (b) Time-domain waveform of the sound pressure signal; (c) Frequency-domain spectrum obtained by FFT.

FIG 7 shows the full cavity evolution and the corresponding acoustic signal for an $L^*$

= 4 projectile at $Fr$ = 9.6. The occurrence of surface-seal closure in this case allows the entire process to be divided into three phases: (A) from the projectile's impact on the free surface to the onset of surface seal; (B) from surface seal to the final cavity pinch-off; and (C) the sustained oscillation stage following pinch-off. The initial behavior of the acoustic signal is similar to that observed in the deep-seal regime. Surface closure occurs at $t_3$ = 38 ms and generates a secondary acoustic pulse, which persists with decaying oscillations until the final cavity pinch-off at $t_4$ = 54.4 ms.

After the longitudinal pinch-off, the cavity separates into two segments while the projectile continues to descend, entrapping a portion of the air. The acoustic pressure signal rises sharply to a peak and then gradually decays, accompanied by bubble shedding from the cavity tail. Compared with the deep-seal case, bubble shedding is more pronounced, resulting in more distinct oscillatory fluctuations after $t_8$.

FIG 7(c) presents the FFT of the acoustic pressure signal shown in FIG 7(b). The dominant frequency of the acoustic signal for this case is approximately 312 Hz. The interval separating the first two positive peaks after $t_4$ yields $\Delta t$ = 3.3 ms, corresponding to a frequency of $f = 1/\Delta t \approx$ 303 Hz. Although the frequency determined from the time-domain signal is slightly lower than the dominant frequency identified in the spectral analysis, the discrepancy is only 2.9%. Moreover, these initial spectral peaks appear prior to the onset of bubble pinch-off, indicating that the dominant oscillation frequency is already established before any significant gas shedding occurs from the cavity. Subsequent gas loss only leads to a slight upward shift in the dominant frequency. Comparative analysis across various *Fr* reveals that an increase in *Fr* yields a larger sealed cavity volume at pinch-off. This larger volume consequently extends the oscillation period, manifesting as a downward shift in the dominant frequency.

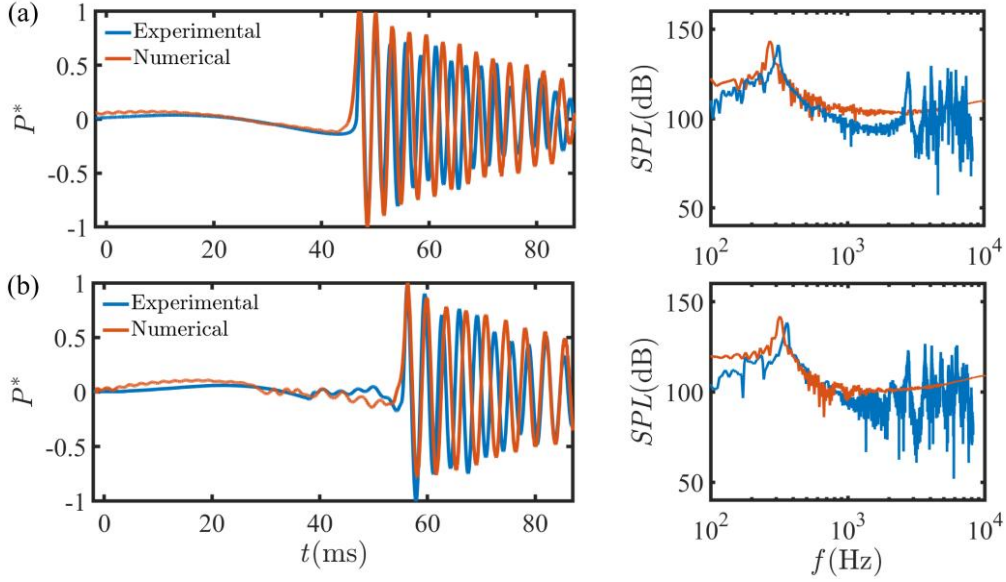

FIG 8. (Color online) Comparison of simulated and experimental acoustic signals at the pressure monitoring point. The left panels present the normalized time-domain waveforms, while the right panels display the corresponding frequency spectra: (a) *Fr* = 8.0; (b) *Fr* = 9.6.

To further investigate the mechanisms responsible for the acoustic signals generated during projectile water entry, numerical simulations are conducted to elucidate the associated acoustic radiation characteristics by analyzing the transient flow-field dynamics. FIG 8 compares the time-domain and frequency-domain acoustic responses obtained from numerical simulations with those measured experimentally at different *Fr*. In the simulations, the sound pressure is defined as the fluctuating component of the relative static pressure, expressed as follows:

$$P'(t) = P(t) - P_0, \qquad (7)$$

Where $P_0$ denotes the static background pressure, which is constant in time and corresponds to the hydrostatic pressure at the measurement location. To emphasize the periodic features and relative amplitudes, the time-domain acoustic pressure is normalized by its maximum value (Sun et al, 2021). The numerical and experimental waveforms exhibit good agreement during the early stage of water entry, but deviations emerge after cavity pinch-off. The phases of the first two positive peaks in the simulation match the experimental measurements, after which a gradual phase lag develops. Moreover, while the simulation predicts a smooth decay, the experimental signal shows distinct non-monotonic fluctuations. These differences stem from inherent limitations of the numerical model in representing key post-pinch-off processes, including cavity-surface instability, rupture, and the associated gas exchange.

In the frequency domain, the simulation fails to capture the multiple high-frequency resonant peaks seen in the experiment. In contrast, it exhibits a certain low-frequency shift in the dominant frequency, with errors of 13.5% and 11.1% for cases (a) and (b), respectively. This discrepancy is primarily attributed to minor differences in the cavity geometry and volume between the numerical model and experimental observations. Furthermore, the irregular gas shedding from the cavity tail in the experiments, while having a limited impact on the overall oscillation period, introduces a slight deviation in the dominant frequency by gradually reducing the effective cavity volume. This shedding also redistributes part of the energy toward higher frequencies, which explains the absence of high-frequency resonant peaks in the numerical results. Despite these quantitative differences, the numerical model captures the overall cavity-closure modes and the key features of the flow-field evolution during water entry and subsequent cavity oscillations, providing a useful numerical reference for interpreting the underlying physics of water-entry cavity dynamics.

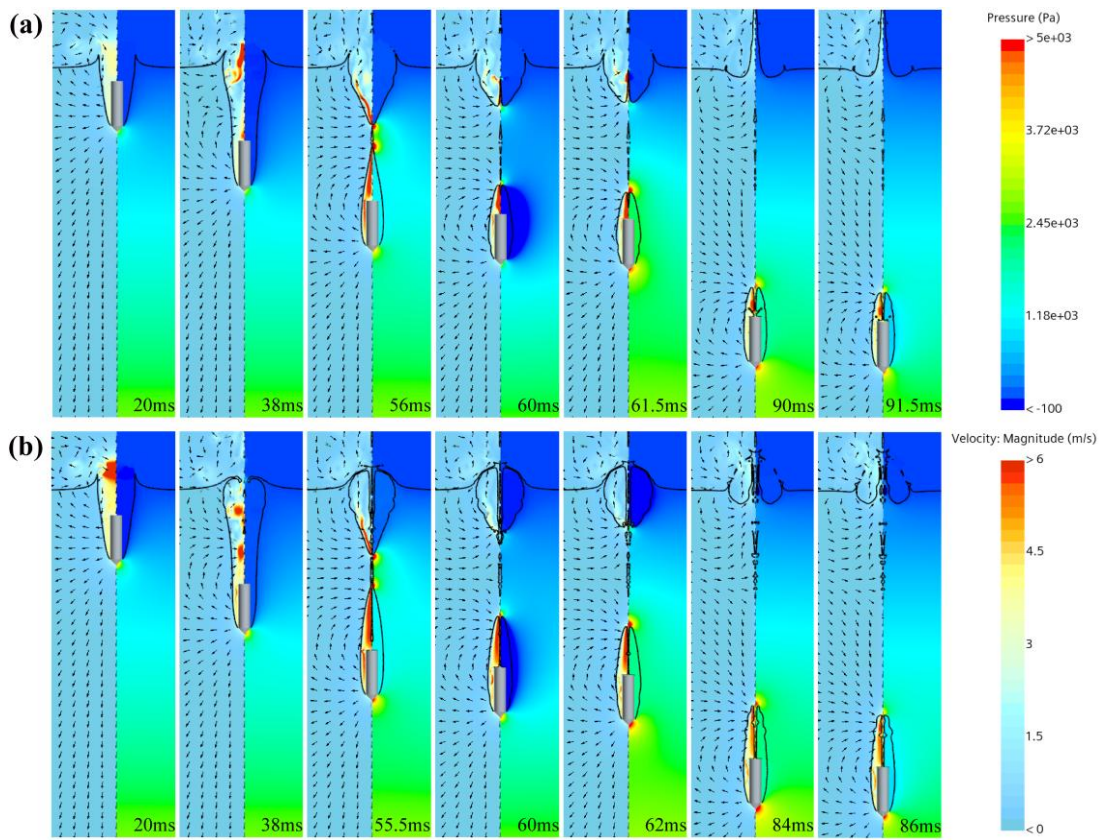

FIG 9. (Color online) Velocity field (left of each panel) and pressure distribution (right of each panel) during the projectile water entry in the numerical simulation. Velocity vectors are normalized to indicate direction only. Color bars represent pressure and velocity magnitude: (a) $Fr$ = 8.0; (b) $Fr$ = 9.6.

FIG 9 presents the velocity field (left) and pressure distribution (right) during water entry for two representative *Fr*. Upon impact, a localized high-pressure zone develops at the projectile tip. As the cavity grows, the internal pressure decreases, drawing air into the cavity and driving the splash crown toward closure. In the high-*Fr* case, surface closure occurs earlier at $t = 38$ ms (FIG 9(b)), preventing further air entrainment. Under both closure modes, the velocity fields show that the projectile imparts radial momentum to the liquid, which reverses the radial flow in the cavity mid-section and initiates the necking that ultimately produces the deep-seal pinch-off ($t = 38$–$55.5$ ms). At pinch-off, flow focusing generates a localized high-pressure region at the neck, exciting volume oscillations within the sealed cavity. These oscillations reflect from the projectile surface and manifest as the visible surface ripples.

The final four frames in FIG 9 show that following pinch-off, the internal cavity pressure undergoes periodic oscillations with a gradually stabilizing period but a decreasing mean pressure, consistent with the pressure time-history in FIG 8. Comparison of the two cases indicates that a higher *Fr* produces a larger sealed cavity volume and consequently yields a longer pressure-oscillation period, as observed in FIG 9(b).

### C. Analysis of the natural frequency of the cavity

Previous studies have shown that the oscillation frequency of a cavity can be affected by confined boundary conditions (Oguz et al., 1998), providing the theoretical foundation for the analysis presented in this section. During water entry, the flow is characterized by a high Reynolds number ($O(10^4)$) and a low Mach number ($Ma < 0.1$). Accordingly, the liquid phase can be approximated as inviscid, irrotational, and incompressible. Moreover, since both the projectile and cavity dimensions are on the centimeter scale, the corresponding Weber number ($We = \varrho U^2 D/\sigma$) is on the order of $O(10^3)$. This high Weber number justifies neglecting surface tension effects, allowing the gas pressure at the cavity interface to be governed primarily by the internal pressure. The present analysis focuses on the sealed cavity formed after underwater pinch-off (e.g., the cavity observed at $t \approx$ 54ms in FIG 3). The normal force balance at the cavity surface is expressed as follows:

$$P_b = P, \tag{8}$$

where $P_b$ denotes the gas pressure inside the cavity. The fluid pressure on the outer surface of the cavity $P$ can be expressed in terms of the velocity potential and the ambient pressure via the Bernoulli equation, with the second-order terms neglected:

$$P = -\varrho \frac{\partial \phi}{\partial t} + P_\infty, \tag{9}$$

where $\varrho$ is the fluid density and $P_\infty$ is treated as a constant. According to Eqs.(8) and (9), the cavity surface is uniform. Assuming the gas pressure inside the cavity is spatially uniform, it follows from Eq.(9) that the internal pressure can be expressed as:

$$P_b = -\varrho \left\langle \frac{\partial \phi}{\partial t} \right\rangle + P_\infty, \tag{10}$$

where $\langle x \rangle$ denotes the mean value of the variable. Neglecting the second-order terms in the material derivative and interchanging the temporal derivative with the surface average, the above equation can be rewritten as:

$$P_b = -\varrho \frac{\mathrm{d}}{\mathrm{d}t} \langle \phi \rangle + P_\infty, \tag{11}$$

taking the time derivative of the above equation yields:

$$\varrho \frac{d^2}{dt^2}\langle\phi\rangle = -\frac{dP_b}{dt}. \tag{12}$$

For linear oscillations at a single angular frequency $\omega$, any variable scales linearly with the others. Given the ideal gas law, which states that the internal bubble pressure $P_b$ is inversely proportional to its volume $V$, the time derivative of the pressure $P_b$ can be expressed as:

$$\frac{dP_b}{dt} = \left(-\frac{dP_b}{dV}\right)\left(-\frac{dV}{dt}\right), \tag{13}$$

where $dP_b/dV$ is a negative value, and $dV/dt$ can be expressed as:

$$-\frac{dV}{dt} = -\int_S \boldsymbol{u}\cdot\boldsymbol{n}\, dS = S\left\langle -\frac{\partial\phi}{\partial n}\right\rangle, \tag{14}$$

where $\boldsymbol{n}$ is the unit normal vector pointing outward from the cavity surface, and $S$ is the surface area of the cavity. Substituting Eqs. (13) and (14) into Eq. (12), and replacing d/dt with $i\omega$ to transform the analysis from the time domain to the frequency domain, Eq. (12) becomes:

$$\omega^2 = -\frac{S}{\varrho\langle\phi\rangle}\frac{dP_b}{dV}\left\langle -\frac{\partial\phi}{\partial n}\right\rangle. \tag{15}$$

The natural frequency of the cavity pulsation is then given by:

$$f = \frac{1}{2\pi}\sqrt{-\frac{S}{\varrho\langle\phi\rangle}\frac{dP_b}{dV}\left\langle -\frac{\partial\phi}{\partial n}\right\rangle}. \tag{16}$$

As mentioned previously, the cavity surface $\phi$ is uniformly distributed. Assuming an unperturbed cavity surface with $\phi=1$ at a given time $t$, the cavity surface area is uniquely determined. The mean normal velocity on the cavity surface can then be solved using the boundary integral method, which has been extensively validated for solving bubble dynamics and fluid–structure interaction problems (Chen et al., 2025; Yi et al., 2021; Li et al., 2018). The rate of change of the internal cavity pressure with respect to volume is obtained by analyzing short-timescale cavity volume variations in combination with the gas state Eq.(6). Based on classical cavity vibration theory, this study further derives an analytical expression for the natural frequency of the cavity under constrained boundary conditions. To further validate the accuracy of this theory, the natural frequency is calculated below using experimental results and specific cavity morphology data. The analysis focuses on the cavity at approximately 54 ms after water entry, when mid-depth pinch-off occurs and a stable air bubble forms attached to the projectile. In this state, key geometric parameters of the cavity profile were extracted, including the boundary depth coordinate $H$ and the radius coordinate $R$.

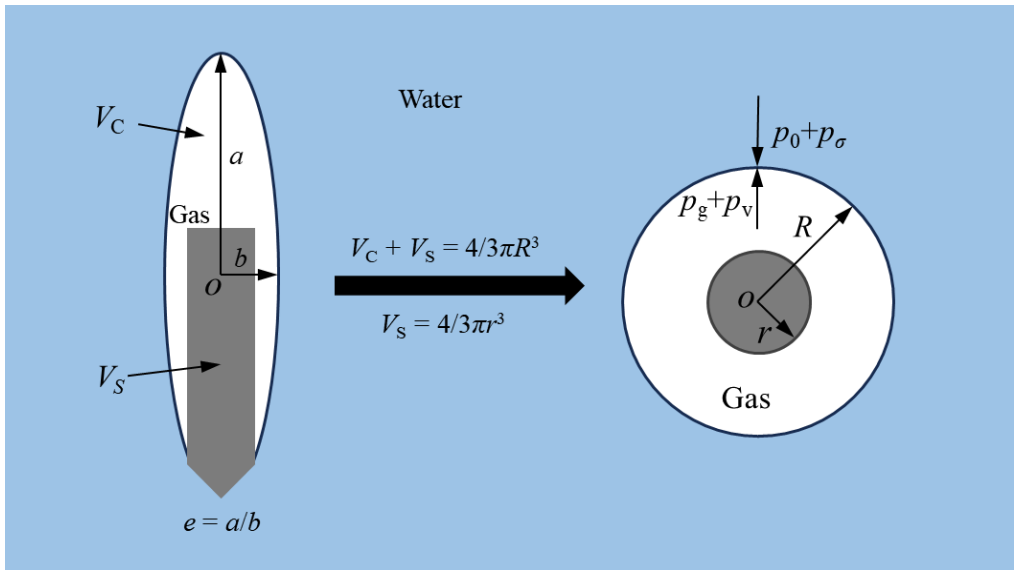

FIG 10. (Color online) A schematic diagram of an ellipsoidal cavity and a spherical cavity with a rigid core and an equivalent volume.

Previous studies have shown that the oscillation frequency of a cavity formed by a sphere entering water closely matches the Minnaert frequency of a spherical bubble of equivalent volume (Grumstrup et al., 2007; Ueda et al., 2022). In contrast, for a conical-nosed projectile, the solid occupies a non-negligible portion of the cavity after pinch-off, and only the finite gas layer surrounding the projectile participates in compression during oscillations. Accordingly, based on the classical Minnaert frequency for a spherical bubble, the effect of the internal projectile is accounted for by modeling the cavity as a spherical bubble containing a rigid core, as illustrated in FIG 10. The left schematic depicts the air cavity remaining attached to the projectile after pinch-off, where $V_C$ and $V_S$ denote the gas-filled cavity volume and the projectile volume, respectively. This configuration is then transformed into an equivalent spherical bubble with the same gas volume, shown in the right schematic. Here, $R$ is the radius of the equivalent sphere corresponding to the combined volume of the cavity and the projectile combined, and $r$ is the radius of the equivalent sphere corresponding to the projectile volume alone. Consequently, the actual cavity gas volume $V_C$ can be expressed as:

$$V_C = V_0 = \frac{4\pi}{3}(R^3 - r^3), \tag{17}$$

assuming the cavity reaches an equilibrium state after pinch-off, the internal gas pressure is given by:

$$P_g = P_0 - P_v + P_\sigma, \tag{18}$$

where $P_0$ is the ambient liquid pressure acting on the outer surface of the bubble, $p_v$ is the vapor pressure acting on the inner surface, $P_\sigma = 2\sigma/R$ and $\sigma$ is the surface tension coefficient. For centimeter-scale projectiles, surface tension effects are negligible.

Consequently, the natural frequency of a spherical bubble containing a rigid core is derived as:

$$\omega_0 = \frac{1}{R}\sqrt{\frac{3\gamma P_0}{\varrho(1-\frac{r^3}{R^3})}}. \tag{19}$$

This shows that the rigid core reduces the effective compressible gas volume. This reduction increases the system's restoring stiffness, which consequently raises the bubble's natural frequency beyond the classical Minnaert prediction. An increase in the rigid core's volume fraction reduces the bubble's compressibility. This drives the oscillation behavior toward that of a rigid body, resulting in a progressively higher frequency. In the limiting case where $r \to 0$, the model reduces asymptotically to the classical Minnaert formulation.

However, cavities formed during water entry typically exhibit significant elongation, which may cause certain sensitivity issues in the definition of the characteristic radius $R$. To refine the model and rigorously account for these geometric effects, we have introduced an equivalent ellipsoidal model based on Strasberg's classical theory (Strasberg, 1953). Strasberg demonstrated that while the stiffness constant (related to the gas compression) remains essentially unchanged for non-spherical bubbles at constant volume, the effective fluid inertia (inertial constant) decreases as the aspect ratio increases. Accordingly, we define an equivalent ellipsoidal cavity with semi-major axis $a$ (half the cavity length) and semi-minor axis $b$ (the radius at the cavity center), with aspect ratio $e = a/b$ (see FIG 10). For the special case of $e = 1$, the geometry reverts to a sphere. Based on Eq. (19), we introduce the mass ratio $m_e / m_0 = e^{2/3} \arctan\sqrt{(e^2-1)} / \sqrt{(e^2-1)}$ to correct the frequency of the equivalent spherical bubble. Since $f_e / f_0 = \sqrt{m_0 / m_e}$, the oscillation frequency of an ellipsoidal bubble with an equivalent volume can therefore be expressed as:

$$\omega_0 = \frac{1}{R}\sqrt{\frac{3\gamma P_0}{\varrho(1-\frac{r^3}{R^3})}}\sqrt{\frac{\sqrt{e^2-1}}{e^{\frac{2}{3}}\arctan(\sqrt{e^2-1})}} \tag{20}$$

To examine the relationship between $Fr$ and $f$ during water entry, FIG 11(a) presents experimentally measured frequencies under different $Fr$ conditions alongside various theoretical predictions. It is evident that the classical Minnaert formulation significantly underpredicts the experimental observations. Introducing an internal rigid core substantially increases the predicted frequency and improves agreement with the measurements, confirming that the presence of an incompressible internal core is a non-negligible factor in determining the cavity oscillation frequency. Building upon this modification, the further incorporation of ellipsoidal corrections leads to an additional, increase in the predicted frequency, indicating that deviations from spherical geometry also play an important role in the oscillation dynamics. Nevertheless, a systematic discrepancy of approximately 10% persists across all theoretical variants, suggesting that these traditional models remain inadequate for describing slender cavities attached to solid surfaces. This limitation arises because the cavity oscillation frequency is strongly governed by both the cavity geometry and the associated boundary conditions, which are only partially represented in simplified analytical theories. By contrast, the boundary integral method inherently accounts for the actual cavity geometry and constrained boundary conditions without requiring ad hoc geometric corrections. The predicted results are in strong agreement with the experimental measurements, exhibiting a maximum deviation below 3%. This close correspondence demonstrates that the cavity oscillation model incorporating constrained boundary conditions effectively captures the oscillation frequency characteristics of projectile-attached cavities.

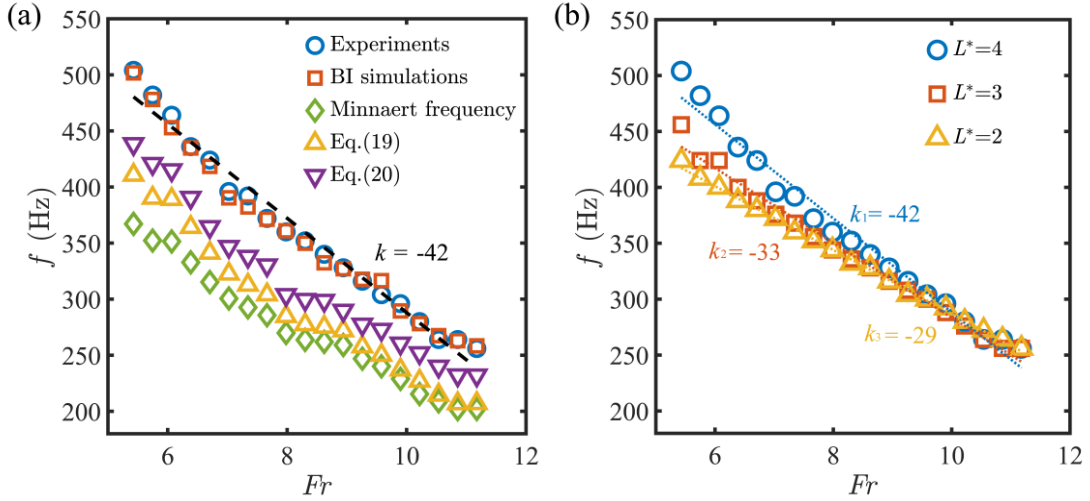

FIG 11. (Color online) Variation of the primary frequency of the cavity with water-entry velocity: (a) Comparison of the cavity oscillation frequency obtained from experiments (blue circles), boundary integral (BI) simulations (red squares), the classical Minnaert frequency (green diamonds), the Minnaert frequency corrected to account for the presence of a rigid core (yellow triangles), and the equivalent ellipsoidal model (purple inverted triangles). The dotted line indicates the linear fit of the experimental results. ($L^* = 4$); (b) Effect of projectile length. The dotted line indicates the linear fit of the experimental results.

A least-squares fit to the experimental data reveals that within the range of $5.4 < Fr < 11.2$, the cavity oscillation frequency follows a linear relationship given by $f = -42Fr + 709$. Notably, when $Fr$ falls within the lower range ($5.4 < Fr < 6.4$), both experimental and model results deviate from this fit. This deviation arises from premature cavity truncation: at lower $Fr$, pinch-off occurs closer to the free surface, and the excessively large aspect ratio $L^*$ of the projectile in experiments causes the cavity to be prematurely truncated along the projectile surface, preventing it from fully enveloping the entire body. Consequently, the actual cavity volume is reduced, leading to an increase in oscillation frequency. These observations confirm that the theoretical model remains applicable under high-Froude-number conditions.

Subsequently, the influence of the projectile aspect ratio $L^*$ on the cavity oscillation frequency $f$ was investigated. FIG 11(b) shows the relationship between $Fr$ and $f$ for three values of $L^*$, indicating a linearly decreasing trend in all cases. The corresponding coefficients $k_1$, $k_2$, and $k_3$ for $L^* = 4$, 3, and 2 are $-42$, $-33$, and $-29$, respectively. At lower $Fr$ values, the frequency $f$ increases with $L^*$, though this difference narrows as $Fr$ increases. Near $Fr \approx 10.9$, the frequencies $f$ for the three $L^*$ values become nearly identical. Beyond this point, as $Fr$ continues to rise, the frequency exhibits a reversal in its dependence on $L^*$, decreasing with increasing aspect ratio.

To clarify the physical mechanism behind this behavior, the following analysis focuses on the velocity evolution of the projectile during water entry. After impact, the dominant decelerating force acting on the projectile is the hydrodynamic drag, given by:

$$F_d = \frac{1}{2} \varrho_{\text{water}} C_d A U^2, \qquad (21)$$

where $\varrho_{\text{water}}$ is the water density, $C_d$ is the drag coefficient, $A$ is the projectile's cross-sectional area, and $U$ is the instantaneous velocity during water entry. For subsonic water entry, $C_d$ is commonly treated as constant across different impact velocities and penetration depths (Abraham et al., 2014). The resulting acceleration $a$ of the projectile is therefore:

$$a = \frac{F_d}{m} = \frac{\frac{1}{2} \varrho_{\text{water}} C_d A U^2}{m} \sim \frac{U^2}{L^*}. \qquad (22)$$

Based on the relationship between acceleration $a$ and aspect ratio $L^*$, projectiles with a larger $L^*$ exhibit greater inertia during water entry, leading to reduced acceleration and slower velocity decay. Under low $Fr$ conditions, where the initial impact velocity is small, the acceleration $a$ is minimal, and the influence of $L^*$ on acceleration $a$ becomes negligible. As a result, the variation in projectile velocity is insignificant. Experimental observations confirm that at the moment of deep-seal pinch-off, the cavities generated by the three projectiles with different $L^*$ values exhibit nearly identical shapes. However, due to the volume occupied by the projectile itself, the actual gas-filled cavity volume differs among the cases. Specifically, a larger $L^*$ results in a smaller gas volume, which in turn raises the oscillation frequency $f$.

As $Fr$ increases, the effect of $L^*$ on acceleration becomes more pronounced. The projectile velocity decays more gradually, enabling the cavity to extend deeper and leading to a larger volume after pinch-off. When $Fr > 10.9$, the increased cavity volume becomes sufficient to compensate for the volume occupied by the internal projectile. In this regime, a larger $L^*$ corresponds to a greater actual cavity volume, which consequently results in a lower oscillation frequency $f$.

## IV. CONCLUSION

We experimentally, numerically and theoretically examine the underwater sound and its origin in cavity collapse during the water entry of conical-nosed projectiles. Particular attention is given to how the Froude number $Fr$ and the aspect ratio $L^*$ control the sealed-cavity oscillation frequency $f$. The numerical model reproduces the key experimental phenomena, and a potential-flow model accurately predicts the natural frequency of the elongated cavity. The main findings are:

The Froude number $Fr$ is identified as the primary controlling parameter for the cavity closure mode. As $Fr$ increases—achieved through higher entry velocities—the internal cavity pressure drop induced by air flow becomes more pronounced, thereby favoring surface seal. Deep seal is driven by hydrostatic pressure, manifesting as cavity contraction and eventual pinch-off near the mid-section. With increasing $L^*$, reflecting greater projectile length and mass, the underwater cavity pinch-off is delayed, and the final sealed cavity volume increases, thereby altering its subsequent acoustic radiation characteristics.

We show that the underwater sound is governed by the sealed-cavity volume pulsation. Higher $Fr$ enlarges the trapped bubble, lengthens the oscillation period and lowers the dominant frequency. Finite-volume simulations reproduce both the pressure decay and the spectral peak. Flow visualisation further reveals that a high-pressure pulse generated at the pinch-off neck provides the impulsive excitation that triggers the cavity volume oscillations.

To quantitatively predict the frequency of cavity-induced acoustic waves, a semi-theoretical framework that explicitly incorporates projectile-boundary effects was developed and validated by the boundary-integral method. In a preliminary improved Minnaert-based model, part of the fluid is replaced by a rigid core and an ellipsoidal correction is applied, thereby reducing the difference between the measured frequency and the actual frequency to approximately 10%. A refined model that fully accounts for the finite projectile surface further collapses the error below 3 %. For the present aspect ratio ($L^* = 4$) the cavity oscillation frequency decreases linearly with the Froude number according to $f\,(\mathrm{Hz}) = -42\,Fr + 709$.

Finally, this study elucidates how projectile geometry governs the acoustic signature. The aspect ratio $L^*$ tunes the oscillation frequency through two competing mechanisms that shift the coefficient $k$ in the linear $f$–$Fr$ relation. At low $Fr$, the volume-occupation effect dominates: a longer projectile displaces more air, shrinking the trapped-gas volume and raising the frequency. At high $Fr$, the inertial effect takes over: the greater projectile mass sustains a deeper, larger cavity, expanding the gas volume and consequently lowering the frequency.

## V. ACKONWLEDGEMENTS

This work is supported by the National Natural Science Foundation of China (Nos. 12372239 and 52525102), the Key R&D Program Project of Heilongjiang Province (JD24A002).

## VI. AUTHOR DECLARATIONS

### Conflict of Interest

The authors have no conflicts to disclose.

## VII. DATA AVAILABILITY

The data that support the findings of this study are available from the corresponding author upon reasonable request.